\documentclass[12pt,preprint]{aastex}

\begin{document}
 
\title{Interferometric Observations of V838 Monocerotis} 
\author{B. F. Lane \altaffilmark{1}, A. Retter \altaffilmark{2}, R. R. Thompson \altaffilmark{3}, J. A. Eisner\altaffilmark{4}} 

\altaffiltext{1}{Center for Space Research, MIT Department of Physics, 70 Vassar Street, Cambridge, MA 02139; blane@mit.edu}
\altaffiltext{2}{Astronomy \& Astrophysics Dept., Penn State University, 525 Davey Lab, University Park, PA 16802-6305; retter@astro.psu.edu}
\altaffiltext{3}{Michelson Science Center, 100-22 California Institute of Technology, Pasadena, CA 91125;thompson@ipac.caltech.edu }
\altaffiltext{4}{Department of Astronomy, California Institute of Technology, MC 105-24, Pasadena, CA 91125; jae@astro.caltech.edu.}

\begin{abstract} 

We have used long-baseline near-IR interferometry to resolve the
peculiar eruptive variable V838 Mon and to provide the first direct
measurement of its angular size. Assuming a uniform disk model for the
emission we derive an apparent angular diameter at the time of
observations (November-December 2004) of $1.83 \pm 0.06$
milli-arcseconds. For a nominal distance of $8\pm2$ kpc, this implies
a linear radius of $1570 \pm 400 R_{\odot}$. However, the data are
somewhat better fit by elliptical disk or binary component models, and
we suggest that the emission may be strongly affected by ejecta from
the outburst.
\end{abstract}


\section{Introduction}

V838 Monocerotis is an eruptive variable star that underwent a
nova-like event in early 2002 \citep{b02,mun02}, with a peak intensity
of $m_V \sim 6.8$.  However, the eruption was unlike classical novae
in that the effective temperature of the object dropped and the
spectral type evolved into a very late M--L type
\citep{evans03}. Another unusual behavior was that the light-curve
displayed multiple peaks in intensity, with the second peak being the
brightest \citep{rm03}. Spectroscopic studies indicated the presence
of $s$-process elements, followed by the appearance of numerous
oxygen-rich molecules \citep{lynch04}.  The eruption mechanism of V838
Mon remains unclear, but appears to represent a new type of explosive
variable. There have been two other variables that displayed similar
late, cool spectral properties: M31-RV \citep{r89,m90,br92} which 
achieved $M_V = -9.95$, and V4332 Sgr (\citet{m99};$M_V \sim -4.5$ though
the distance to V4332 Sgr is uncertain).

The discovery of a spectacular light-echo effect around V838 Mon
\citep{hend02} has allowed distance estimates to be made
\citep{wis03,bond03,tyl04,vl04}; they indicate distances in the range
2.5 -- 10 kpc, depending on the location of the reflecting material
(circumstellar vs. interstellar.) It has been pointed out
\citep{tyl05} that the apparent center of the light echo appears to
move on the sky; this is inconsistent with a circumstellar dust model
and would appear to favor the interstellar interpretation.  Assuming a
distance of $8 \pm 2$ kpc \citep{tyl04}, the peak luminosity was $\sim
6.4 \times 10^5 L_{\odot}$ and the absolute visual magnitude was in
the range $M_V \sim -7.7$ to $-10$. As of late 2004 the source had
dimmed considerably in the visible from peak intensity ($m_V = 14.82$;
Kiss, personal communication to A.R.), but remained bright in the
near-IR ($m_K = 5.52$; Ashok \& Banerjee, 2004).

The progenitor of V838 Mon has been found
\citep{mudes02,ws02,kim02,gor04} in a variety of surveys, including
2MASS \citep{2mass}, and appears consistent with a somewhat reddened
early main sequence star (B3V): $B=15.8 \pm 0.06$,~$K=13.33 \pm
0.06$,~$B-V \sim \pm 0.6$,~$E_{B-V}\sim 0.77$. During the 66 years
preceding the 2002 outburst the progenitor was not significantly
variable \citep{gor04}. \citet{kato02} point out that the progenitor
was detected by IRAS (IRAS 07015-0346). Post-outburst observations
have found a faint, blue component in the spectrum
\citep{desmun02,ws02} and \citet{tyl05} argued that the photometry is
well-matched by a pair of early main sequence stars (B3V + B1.5V or
B4V + A0.5V). \citet{lynch04} observed the IR behavior of V838 Mon and
fit their data to a model consisting of a cool ($T_p=2100$K,
$R_p=8.8$AU) stellar photosphere surrounded by a large, absorbing
molecular cloud ($T=850$K, $R_c=43$AU), deriving a total mass of the
ejecta of 0.03--0.13$M_{\odot}$.

Several models have been proposed to explain the outburst of V838
Mon. Initial models included thermonuclear runaway on an old, cool
white dwarf \citep{tut92}, though this may be hard to reconcile with
the presence of a B3V companion. Another possibility is the merger of
a pair of main-sequence binary stars \citep{sk03}.  It has also been
suggested that V838 Mon is a post-AGB star and the event was due to
He-flash \citep{vl04}. Finally, \cite{rm03} have suggested that the
eruption was due to the accretion of several close-in giant planets by
an expanding giant or AGB star. This is consistent with the appearance
of multiple peaks in the light curve, and with the spectroscopic
detection of lithium \citep{mun02}.

We have used the Palomar Testbed Interferometer (PTI) to resolve the
$2.2 \mu$m emission from V838 Mon and measure its apparent angular
diameter. The Palomar Testbed Interferometer (PTI) was built by
NASA/JPL as a testbed for developing ground-based interferometry and
is located on Palomar Mountain near San Diego, CA \citep{col99}. It
combines starlight from two 40-cm apertures and measures the resulting
interference fringes. The high angular resolution provided by this
long-baseline (85-110 m), near infrared ($2.2 \mu$m) interferometer is
sufficient to resolve emission on the milli-arcsecond scale.  The measured 
apparent angular diameter can be combined with distance estimates from
the light-echo to determine the size of the emitting region and thus
help constrain explosion models.

\section{Observations}

We observed V838 Mon on 6 nights between 5 November and 13 December
2004 (mean modified Julian Date=53338.3), using PTI in the standard
K-band mode, with the 85-meter North-West (4 nights) and South-West (2
nights) baselines. For detailed descriptions of the instrument we
refer the reader to \cite{col99}.  Each nightly observation consisted
of one or more 130-second integrations during which the normalized
fringe visibility (or contrast) of the science target was
measured. The measured fringe visibilities of the science target were
calibrated by dividing them by the instrument point-source response of
the instrument (typically $\sim 0.75$), determined by interleaving
observations of calibration sources (Table \ref{tab:calibs}); the
calibration sources were chosen to be single stars with angular
diameters smaller than 1 milli-arcsecond, determined by fitting a
black-body to archival broadband photometry. While HD 49933 is listed
as a double star in the SIMBAD database, the companion is faint ($m_V
= 11.3$) and distant ($\sim 6$ arcseconds) from HD 49933, giving it a
negligible impact on the measured fringe visibilities.  For further
details of the data-reduction process, see \citet{colavita99b} and
\citet{boden00}. Note that in addition to the uncertainties derived
from the internal scatter in the data, we add a 5\% minimum systematic
uncertainty to the visibility measurements, consistent with worst-case
estimates for PTI data based on comparisons with known sources.

PTI is equipped with a low-resolution spectrometer which provides
fringe visibility measurements and photon count rates in five spectral
channels across the K band. We compute a wide bandwidth average as the
photon-weighted average of the five spectral channels. For the
observations of V838 Mon, the photon rates in the two edge channels
(2.0 and 2.4 $\mu$m) were too low to provide useful data, and thus the
fringe visibilities are effectively measured from 2.1 to 2.3
$\mu$m. The faintness of the source in the edges of the K-band is
consistent with the deep molecular absorbtion bands seen by
\citet{ba02} using near-IR spectroscopy. Using photon count rates from
PTI, and K-band magnitudes for the calibrator sources provided by
2MASS \citep{2mass}, we derive a K-band apparent magnitude of $5.45
\pm 0.1$ for V838 Mon, consistent with the measurement by \citet{ab04}
of $m_K = 5.52 \pm 0.05$ made on 4 October 2004. We also note that
within the limit of our statistical measurement noise ($\sim 0.06$
mag) the apparent magnitude did not vary during the period of
observations.

\section{Results}

\begin{figure}[h]
\epsscale{1.0}
\plottwo{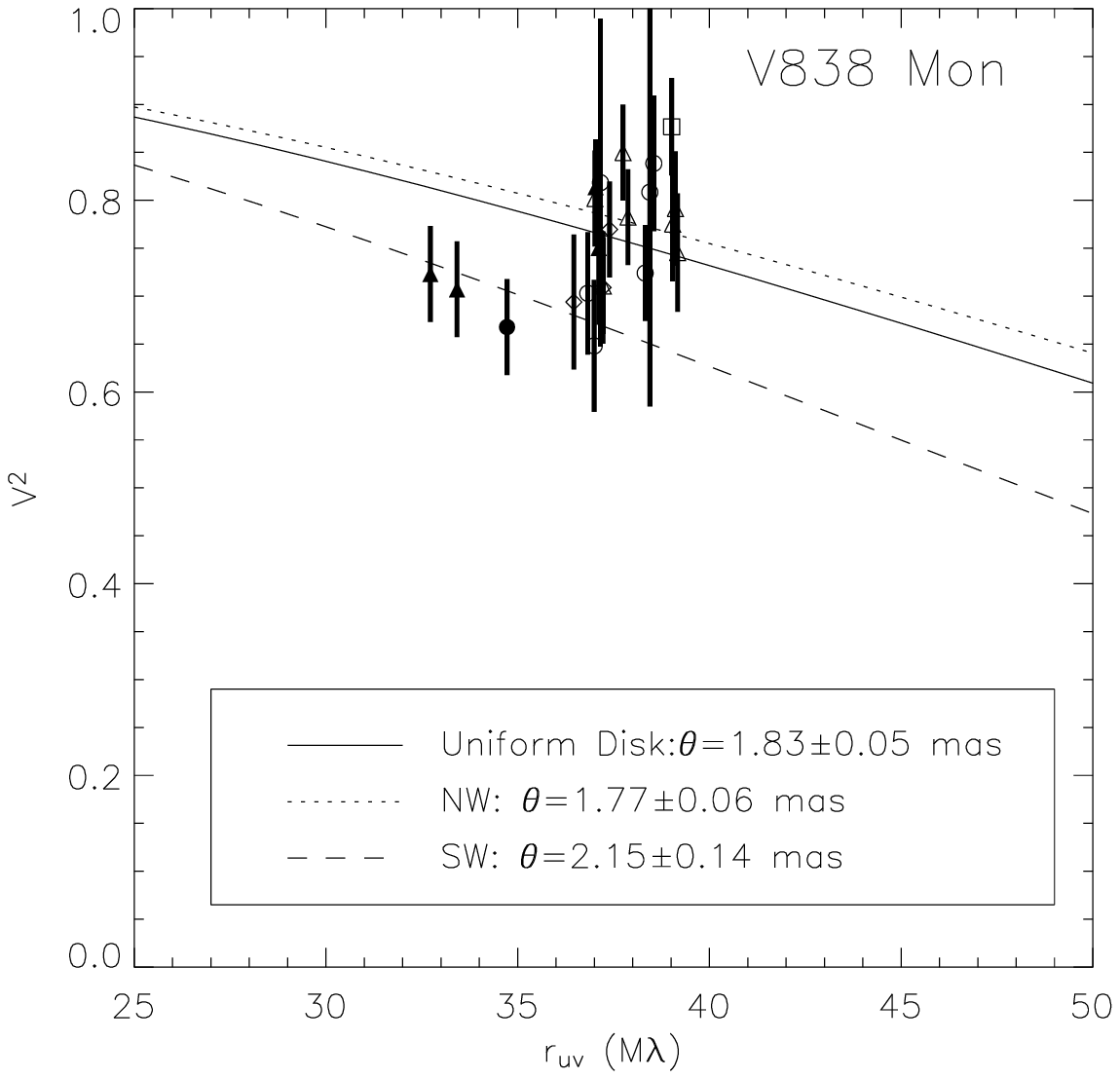}{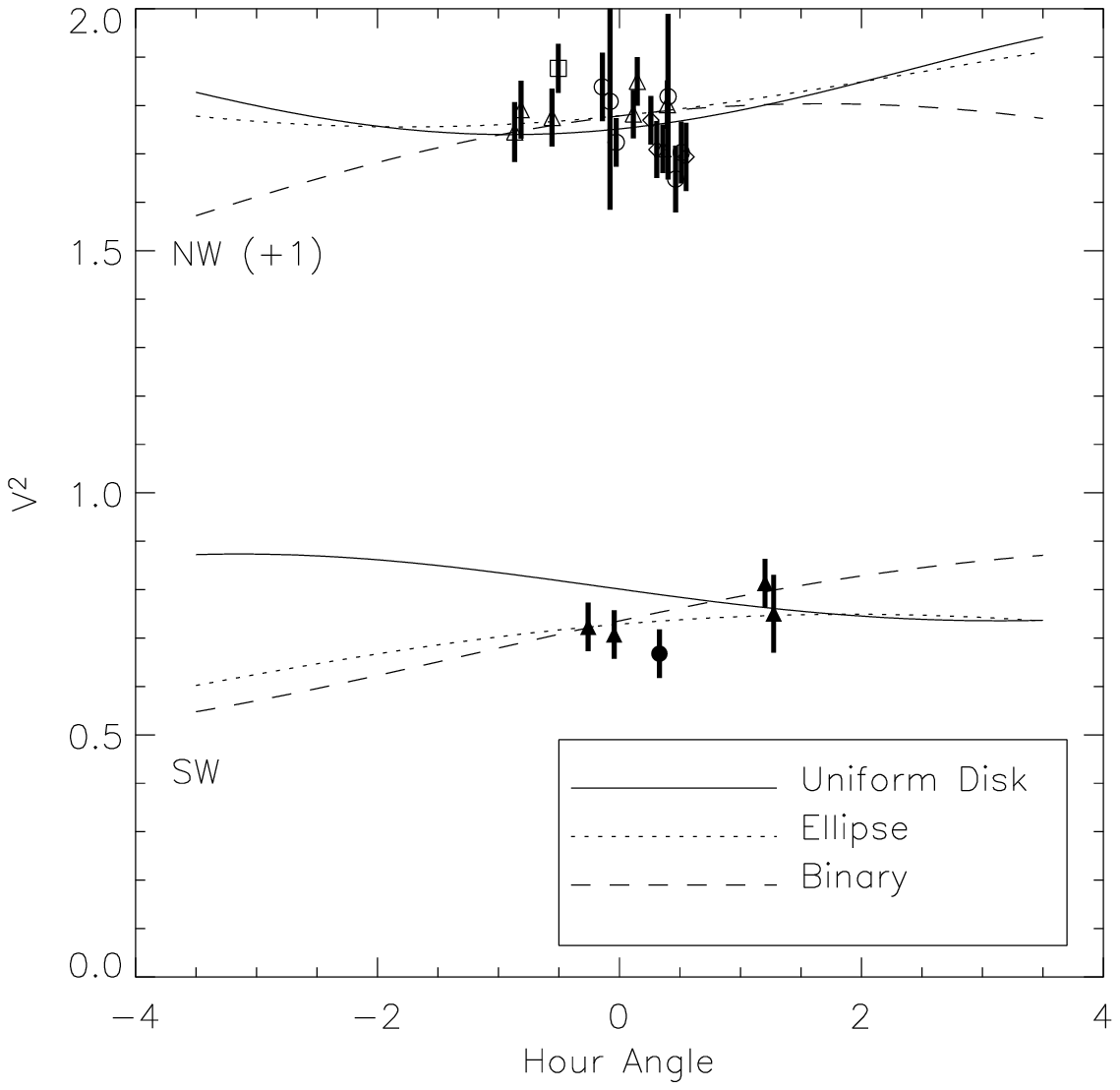}
\caption[]{\label{fig:data} Left: The measured fringe visibility
($V^2$) of V838 Mon as a function of projected baseline length
measured in units of the observing wavelength ($2.2 \mu$m).  Different
symbols are used for different nights. Filled symbols are used for
data taken with the South-West baseline, open symbols indicate
North-West baseline data. Best-fit uniform-disk models are shown for
combined and single-baseline data sets. The data show that the
emission is resolved and is inconsistent with an circularly symmetric
emission source. Right: The measured fringe contrast as a function of
source hour angle, together with the best-fit models. A binary or
inclined disk model is required to account for the data from both
baselines. Note that the NW-baseline data have been moved up by 1.0
for clarity.}
\end{figure}

The theoretical relation between source brightness distribution and
fringe visibility is given by the van Cittert-Zerneke theorem, with
the two being a Fourier-transform pair. For a uniform intensity
disk model the normalized fringe visibility can be related
to the apparent angular diameter using
\begin{equation}
V^2 = \left( \frac{2 \; J_{1}(\pi B \theta_{UD} / \lambda)}{\pi B\theta_{UD} / \lambda} \right)^2
\label{eq:V2_single}
\end{equation}
where $J_{1}$ is the first-order Bessel function, $B$ is the projected
aperture separation, $\theta_{UD}$ is the apparent angular diameter of
the star in the uniform-disk model, and $\lambda$ is the wavelength of
the observation. It is not necessarily true that the source is
well-modeled by a uniform disk; it may be better modeled by a
limb-darkened disk or even a Gaussian extended envelope
\citep{perrin04}.  However, there is insufficient data at this point
to make meaningful comparisons between Gaussian and uniform disk
models, and our uniform disk fits should be interpreted as measuring
the angular extent of the source without providing detailed
information on the radial intensity profile. On the other hand, the
availability of data taken with two different baselines (and hence two
different source position angles) provides information about the
angular intensity profile, and in particular we are able to
distinguish between a circularly symmetric source and elliptical
models (see \citet{eis03} for a discussion of simple emission models).
As indicated in Figure \ref{fig:data} the NW and SW data sets give
differing apparent diameters, indicating that the source is not
circularly symmetric. We fit the combined (NW+SW) data set using an
elliptical model and solve for major and minor-axis diameters and
position angle. Another possible emission morphology is a pair of
sources; we model such a binary source using standard expressions
\citep{eis03} and fit for the binary separation, position angle and
intensity ratio.

 We performed least-squares fits of uniform disk, elliptical and
binary models to the measured fringe visibilities. The results are
given in Table \ref{tab:fits} and Figure \ref{fig:data}. We find that
the circularly-symmetric uniform disk model does not match the data
particuarly well ($\chi^2_{d.o.f} = 1.7$, where $\chi^2_{d.o.f}$ is
the sum of the squares of the residuals, divided by the number of
degrees of freedom). There is insufficient data to distinguish between
an elliptical ($\chi^2_{d.o.f} \sim 1.2$) or a binary emission model
($\chi^2_{d.o.f} \sim 0.92$). It is likely that an even more
complicated morphology will ultimately be required to fully match the
data. 

We examine the possibility of systematic errors mimicing the effect of
non-symmetric emission using the following test: we use one of the
calibrators (HD 49434) as a ``check star'', calibrating it using our
second calibrator (HD 49933), and fitting a circularly-symmetric
uniform disk model (expected to be a good fit for a single
main-sequence star) to this data set. The fit has $\chi^2_{d.o.f} =
1.4$ for 14 points (including points measured on both available
baselines), and produces an apparent angular diameter of $0.39 \pm
0.25$ milli-arcseconds (data not shown), fully consistent with the
expected diameter based on spectrophotometry. The large uncertainty in
the final diameter is merely due to the fact that such a small source
is very neary indistinguishable from a point source using the
available baselines.

We would not expect to see the putative B3V binary companion to V838
Mon, as it would have a contrast ratio in the K band of $\sim 10$
magnitudes. Such a large magnitude difference would
modulate the apparent fringe visibility by at most $10^{-8}$, much
smaller than our measurement precision.

\section{Discussion}

We have modeled the $2.2 \mu$m emission from V838 Mon using several
simple emission morphologies, including uniform and elliptical disks
and binary models. We find that the best fits to the data are provided
by elliptical or binary models. However, the preference is not
overwhelming. We also note that it is quite likely that the emission
morphology is more complicated than these simple models indicate;
further observations are certainly required. We caution the reader
that the stated parameter uncertainties are merely indications of the
statistical uncertainties. There is a larger, and less
well-quantified, uncertainty associated with the choice of emission
model.

If the emission is interpreted as coming from a circularly-symmetric
uniform disk, the best-fit size is $ 1.83 \pm 0.06$
milli-arcseconds. For a distance of $8\pm2$ kpc, derived from the
light-echo, the implied linear radius is $7.3 \pm 1.8$ AU ($1570 \pm
400 R_{\odot}$.)  Such a large stellar radius is consistent with
previous estimates made during the outburst based on fitting an
emission model to spectro-photometry: \citet{lynch04} derive a stellar
photosphere radius of 5.6 AU, assuming a distance of $6$ kpc. Scaling
that result to a distance of 8 kpc would imply an apparent radius of
7.5 AU. However, \citet{tyl05b} derives an apparent radius at the time
of our observations of $\sim 800 R_{\odot}$, somewhat inconsistent
with this result. In any case, such a large radius cannot be the
radius of the white dwarf. In a nova outburst the system returns to
its normal dimension in a few weeks or months - much less than the 3
years elapsed before our observations \citep{h97,r97,r99}.

It is interesting to compare the measured apparent radius with what
might be expected based on simple photometric models: if we fit a
black-body model to the observed near-contemporaneous JHK magnitudes
(October 2004; Ashok \& Banerjee 2004) we derive a temperature of $T =
1950 \pm 100$K, a bolometric luminosity of $\sim 30000 L_{\odot}$ and
an apparent angular diameter of $\theta = 1.3 \pm 0.6$
milli-arcseconds, marginally consistent with our measured
value. However, a black-body is not a particularly suitable model for
a 2000K stellar atmosphere. A better approximation can be provided by
empirical surface-brightness relations. Using the same photometry, and
assuming a reddening correction $A_V = 2.3$ together with V- and
K-band surface-brightness relations from \citet{g04} the available
photometry predicts apparent angular diameters in the range $\sim
0.6-0.7$ milli-arcseconds. However, it should be noted that the $J-K$
color ($\sim 1.5$) is somewhat out of the range of colors used to
derive the relation (-0.2--1.3). 

The best formal fit to the data is given by a binary source
model. However, this result must be interpreted with caution. The
intensity ratio between the components, though uncertain, is nearly
equal ($R = 0.43^{+0.57}_{-0.34}$) in the K-band. Given that the
K-band emission is dominated by a single cool blackbody that appeared
after the outburst, it is hard to imagine a scenario where the two
sources are not ultimately the result of a single outburst
mechanism. The best-fit separation of the binary components is
$5.51^{+5.5}_{-0.13}$ milli-arcseconds ($44.1^{+42}_{-1}$
AU). Another possibility indicated by the data is that the observed
K-band emission is produced by a very elongated, elliptical
structure. The projected linear dimensions of such a source are
approximately $3.5^{+0.2}_{-1.5} \times 0.07^{+3.0}_{-0.07}$
milli-arcseconds ($28^{+2}_{-13} \times 0.5^{+24}_{-0.6}$ AU).

We suggest that the observed K-band emission may be in part due to
ejecta produced during the eruption. Soon after the peak of the
outburst, the expansion velocity of the ejecta was estimated as 50-350
km/s \citep{mun02,c03,kip04}. For the time elapsed since outburst of
$\sim 10^8$ s, the distance between the ejecta and the star should be
in the range 30--220 AU. This range is consistent with the projected
separations we measure. It is also likely that the ejecta are not
uniformly distributed, and hence the binary morphology preferred by
the data may be ``clumpiness'' in the ejecta. In this context we note
that polarimetric observations during the outburst \citep{wis03,des04}
indicate that the source became significantly polarized (0.7\% at
5000\AA). This has been interpreted as being due to departures from
spherical symmetry during the outburst.

\section{Conclusion}

We have used the Palomar Testbed Interferometer to observe the
peculiar outbursting variable V838 Mon. This is only the second
published observation of a nova-like outburst using an optical/near-IR
interferometer - the only previous result being the observations of
Nova Cyg 1992 done using the Mk III interferometer \citep{q93}. The
high angular resolution provided by PTI allows us to draw two
preliminary conclusions. First, the source is resolved by our
instrument and has a characteristic angular size of a few
milli-arcseconds. Second, the difference in measured fringe visibility
between the two baselines is inconsistent with the source being a
simple uniform disk.  We have explored other possible source
morphologies, including inclined disks and binary models and find that
the data can be reasonably matched by such models, but clearly further
data are needed to constrain them.  We point out that any of the
models proposed to explain the outburst of V838 Mon need to account
for the presence of very extended, cool, non-symmetric emission. 

\acknowledgements We wish to acknowledge the extraordinary efforts of
K. Rykoski, and to thank an anonymous referee for helpful
comments. Observations with PTI are made possible through the efforts
of the PTI Collaboration, which we acknowledge. This research has made
use of the Michelson Science Center at Caltech
(http://msc.caltech.edu), JPL/NASA, the Simbad database CDS
(Strasbourg, France), and of data products from the Two Micron All Sky
Survey, which is a joint project of the University of Massachusetts
and the IPAC/Caltech, funded by NASA and NSF. BFL acknowledges support
from a Pappalardo Fellowship in Physics and JAE is grateful for a
Michelson Graduate Fellowship.

\clearpage

\begin{table}
\begin{center}
\begin{tabular}{lccccc}
           &       &   &             &                     &  \\
\tableline
\tableline
Calibrator & V     & K & Spectral    & Ang. Diameter       &    Separation \\
           &       &   & Type        & $\theta_{UD}$ (mas) &     (deg) \\
\tableline
HD 49434   & 5.75  & $5.01$    &  F1V        & $0.36 \pm 0.1$      &    4.7 \\
HD 49933   & 5.78  & $4.72$    &  F2V        & $0.42 \pm 0.1$      &    4.7 \\

\tableline
\end{tabular}
\caption{\label{tab:calibs} Relevant parameters of the calibrators. The 
separation listed is the angular distance from the calibrator to V838 Mon. }
\end{center}
\end{table}
\clearpage

\begin{table}
\begin{center}
\begin{tabular}{lccccc}
            &  &   &          &      &                \\
\tableline
\tableline
Date & Epoch   & Baseline  &  $u$       &  $v$   & Cal. Visibility \\
     & (MJD)  & &  (m)     &  (m) &  ($V^2$)        \\
\tableline
11/5/2004 & 53314.488 & NW & -82.3009 & -20.7979 & $    0.838 \pm 0.071$  \\
11/5/2004 & 53314.491 & NW & -82.0364 & -20.7053 & $    0.808 \pm 0.223$  \\
11/5/2004 & 53314.493 & NW & -81.8108 & -20.6322 & $    0.724 \pm 0.050$  \\
11/5/2004 & 53314.511 & NW & -79.3348 & -20.0247 & $    0.818 \pm 0.171$  \\
11/5/2004 & 53314.513 & NW & -78.8844 & -19.9368 & $    0.648 \pm 0.069$  \\
11/5/2004 & 53314.515 & NW & -78.5433 & -19.8732 & $    0.703 \pm 0.064$  \\
12/3/2004 & 53342.382 & NW & -83.6646 & -21.8587 & $    0.745 \pm 0.062$  \\
12/3/2004 & 53342.384 & NW & -83.6667 & -21.7817 & $    0.791 \pm 0.060$  \\
12/3/2004 & 53342.395 & NW & -83.4496 & -21.4042 & $    0.775 \pm 0.060$  \\
12/3/2004 & 53342.422 & NW & -81.1080 & -20.4302 & $    0.782 \pm 0.050$  \\
12/3/2004 & 53342.424 & NW & -80.9275 & -20.3832 & $    0.850 \pm 0.050$  \\
12/3/2004 & 53342.433 & NW & -79.6297 & -20.0848 & $    0.711 \pm 0.050$  \\
12/3/2004 & 53342.434 & NW & -79.3825 & -20.0341 & $    0.802 \pm 0.050$  \\
12/4/2004 & 53343.394 & NW & -83.3603 & -21.3288 & $    0.877 \pm 0.051$  \\
12/11/2004 & 53350.410 & SW & -47.9772 &  59.9114 & $    0.668 \pm 0.050$  \\
12/12/2004 & 53351.404 & NW & -80.2693 & -20.2247 & $    0.770 \pm 0.050$  \\
12/12/2004 & 53351.406 & NW & -79.9625 & -20.1562 & $    0.709 \pm 0.059$  \\
12/12/2004 & 53351.416 & NW & -78.2418 & -19.8191 & $    0.694 \pm 0.070$  \\
12/13/2004 & 53352.380 & SW & -41.6662 &  59.4457 & $    0.723 \pm 0.050$  \\
12/13/2004 & 53352.389 & SW & -44.0813 &  59.6069 & $    0.707 \pm 0.050$  \\
12/13/2004 & 53352.440 & SW & -55.2141 &  60.7062 & $    0.814 \pm 0.050$  \\
12/13/2004 & 53352.443 & SW & -55.6835 &  60.7759 & $    0.750 \pm 0.080$  \\
\tableline
\end{tabular}
\caption{\label{tab:data} Calibrated fringe visibilities of
V838 Mon, together with the projected baseline components.
The effective wavelength of observations was 2.2 $\mu$m.
 $u,v$ are the components of the projected baseline.}
\end{center}
\end{table}

\clearpage

\clearpage

\begin{table}
\begin{center}
\begin{tabular}{lccccc}
                 &            &                       &                &             \\
\tableline
\tableline
Model            &  $\chi_r^2 $&   Size                & Pos. Angle      & Minor axis \\
                 &             &   (mas)               &  (deg)          & or intensity ratio.      \\
\tableline
Uniform Disk	 &	1.7 &	$1.83^{+0.06}_{-0.06}$&	---	        &	---\\
Elliptical       &	1.3 &	$3.57^{+0.22}_{-1.56}$&	$15^{+3}_{-27}$ & $ 0.07^{+3.03}_{-0.07}$\\
Binary Model	 &	0.9 &	$5.51^{+5.2}_{-0.13} $&	$36^{+3}_{-27}$ & $ 0.43^{+0.57}_{-0.34}$\\
\tableline
\end{tabular}
\caption{\label{tab:fits} Fits of various emission models for V838
Mon. Size refers to the characteristic angular variable: angular
diameter for a uniform disk, major-axis diameter for the elliptical model, and separation
angle for the binary model. For the case of a binary model, the last
column lists the component intensity ratio, $R$, while for the elliptical
model the minor-axis diameter is given. The errors have
been scaled by $\sqrt{\chi^2_{d.o.f.}}$. Note that the position angle has a 
180-degree ambiguity.}
\end{center}
\end{table}

\end{document}